\documentclass[11pt]{article} 
\usepackage{mystyle-new}
\usepackage{authblk}
\usepackage[T1]{fontenc}
\usepackage{epsfig,amsmath} 
\usepackage{hyperref}
\usepackage{color}
\usepackage{graphicx}
\usepackage[utf8]{inputenc}
\usepackage{lmodern}
\usepackage{orcidlink}
\usepackage{etoolbox}     
\usepackage{orcidlink}
\usepackage{xspace}
\definecolor{red}{rgb}{1,0,0}
\def\lesssim{\ \hbox{\raise 2pt \hbox{$<$} \kern -13pt
                     \lower 3pt \hbox{$\sim$}}\ }
\def\greatersim{\ \hbox{\raise 2pt \hbox{$>$} \kern -13pt
                     \lower 3pt \hbox{$\sim$}}\ }

\def\lsim{\mathrel{\rlap{\lower4pt\hbox{\hskip1pt$\sim$}}
    \raise1pt\hbox{$<$}}}                
\def\gsim{\mathrel{\rlap{\lower4pt\hbox{\hskip1pt$\sim$}}
    \raise1pt\hbox{$>$}}}                

\def\rapgap{{\sc RapGap}}
\def\cascade{{\sc Cascade}}
\def\powheg{{POWHEG}}

\input epsf.tex
\def\desepsf(#1 width #2){\epsfxsize=#2 \epsfbox{#1}}
\def\kt{\ensuremath{k_{\rm T}}}

\def\pt{\ensuremath{p_{\rm T}}}

\def\qt{\ensuremath{q_{\rm t}}}

\def\zM{\ensuremath{z_{\rm M}}}

\newcommand{\as}{\ensuremath{\alpha_s}}

\newcommand{\PB}{PB}
\newcommand{\PBset}{{PB-NLO-2018}}

\definecolor{kkcolor}{rgb}{1,0,0}

\newcommand\kkout{\marginpar{\color{kkcolor}$\int$}%
  \bgroup\markoverwith{\color{kkcolor}{\rule[0.4ex]{2pt}{0.8pt}}}\ULon}

\def\pythia{{\scshape Pythia}8\xspace}
\def\pythiaPB{{\scshape{Pdf2Isr}}\xspace}

\newcommand{\GeV}{\text{GeV}}
\newcommand{\TeV}{\text{TeV}}

\def\PZ{\ensuremath{{\rm Z}}}

\def\Pp{\ensuremath{{\rm p}}}

\newcommand{\nnloSplit}{vanNeerven:2000wp,Moch:2004pa,Vogt:2004mw,Vermaseren:2005qc,Blumlein:2021enk,Blumlein:2022gpp,ABLINGER2014263,Ablinger:2017tan,Moch:2014sna,Behring:2019tus,Blumlein:2021ryt}

\newenvironment{tolerant}[1]{\par\tolerance=#1\relax}{ \par }
\usepackage{amsmath,bm}
\usepackage{lineno}

\usepackage{cite,./mcite}
\usepackage{tikz}
\usepackage[symbol]{footmisc}

\newcommand{\ccfm}{Ciafaloni:1987ur,Catani:1989yc,Catani:1989sg,Marchesini:1994wr}

\providecommand{\DOI}[1]{\href{http://dx.doi.org/#1}}

\begin{document}

\title{Entanglement entropy, Monte Carlo event generators, and soft gluons DIScovery} 

\author[1]{M.~Hentschinski\orcidlink{0000-0003-2922-7308}}
\affil[1]{Departamento de Actuaria, Física y Matemáticas, Universidad de las Américas Puebla,
San Andrés Cholula, Puebla, Mexico}

\author[2,3]{H.~Jung\orcidlink{0000-0002-2964-9845}}
\affil[2]{Elementary Particle Physics, University of Antwerp, Belgium}
\affil[3]{II. Institut f\"ur Theoretische Physik, Universit\"at Hamburg,  Hamburg, Germany}
\author[4,5]{K.~Kutak\orcidlink{0000-0003-1924-7372}}
\affil[4]{Institute of Nuclear Physics, Polish Academy of Sciences,  Cracow, Poland}
\affil[5]{CPHT, CNRS, Ecole Polytechnique, Institut Polytechnique de Paris, 91120 Palaiseau, France}

\date{}
\begin{titlepage} 
\maketitle
\vspace*{-18cm}
\begin{flushright}
\end{flushright}
\end{titlepage}

\begin{abstract}
We study entropy production in Deep Inelastic Scattering using Monte Carlo simulations.
We show that the dominant contribution to entropy is due to soft gluons. This contribution is usually neglected in standard Monte Carlo approaches, since it does not affect hadronic spectra. However, it is relevant for entropy and multiplicity distributions, as we demonstrate with explicit calculations. 
We further show that as one includes soft gluons, 
making the Monte Carlo parton distributions closer to inclusive PDFs, the resulting entropy starts to grow with decreasing x. This provides further evidence that the bulk of the measured entropy originates from initial-state effects.
\end{abstract} 

\section {Introduction}
In recent years, questions that have been predominantly explored within Quantum Mechanics, i.e. about manifestation of correlations due to entanglement, are now being investigated in the realm of collider physics. In particular, various authors propose to test Bell inequalities in top decay and spin correlations \cite{Maltoni:2024tul,Maltoni:2024csn,Afik:2025ejh,Hatta:2024lbw,Bhattacharya:2024sno,Altomonte:2024upf,Zhang:2025mmm,Han:2025ewp,Caputa:2024xkp,Qi:2025onf}.
Another line of research is to investigate to what extent entanglement manifests itself in Quantum Chromodynamics (QCD). There are various proposals and many of them are centered on entropy production \cite{Kutak:2011rb,Kutak:2023cwg,Peschanski:2012cw,Stoffers:2012mn,Dumitru:2023fih,Kovner:2015hga,Berges:2017hne,Kovner:2018rbf,Peschanski:2019yah,Dvali:2021ooc,Kutak:2025syp}.

In particular, in Ref.~\cite{Kharzeev:2017qzs} Khazeev and Levin argued that hadronic entropy observed in deep-inelastic scattering (DIS) reactions is entanglement entropy, which is generated during the interaction of the virtual photon and the proton. DIS at low $x$ allows then to test this  proposal since in this limit, the number of produced hadrons is large (and hadronic entropy is therefore sizable), while an approximate determination of the proton wave function is possible through identifying color dipoles as  the natural degrees of freedom in DIS at low $x$.

A natural question to ask in this context is whether  instead of color dipoles one might also consider probability distributions related to the number of partons to determine hadronic entropy, which are naturally provided by the initial state parton shower of  Monte Carlo event generators (MC).  This possibility has been first explored in Ref. \cite{Tu:2019ouv}  before the experimental determination of hadronic entropy by the H1 collaboration~\cite{H1:2020zpd}.  The authors found that entanglement entropy (which they determined approximately through $S = \ln xg$,  with $g(x)$ the gluon distribution function) and the entropy determined from the {\sc Pythia6}~\cite{Sjostrand:2006za}, \pythia \cite{Sjostrand:2014zea,Bierlich:2022pfr} and DJANGO~\cite{Schuler:1991yg} MC event generators show very different behavior. In particular, these MC event generators (studied in Ref. \cite{Tu:2019ouv}) could not reproduce the rise of entropy with decreasing $x$, found in the relation  $S = \ln xG$. However, such a rise of entropy was later on also observed in H1 measurements~\cite{H1:2020zpd}, and these data themselves are very well described by the MC generator \rapgap~\cite{RAPGAP33,Jung:1993gf}.

The current situation is therefore somewhat confusing. To clarify the issue of Monte Carlo event generators,  we perform a detailed study of parton and particle production in DIS using MC event generators.
We find that  correlations, and as such entropy, are first of all due to soft gluons, which are in most Monte Carlo event generators neglected, since they do not play a significant role in the observed hadron spectra.  Soft gluons are however important to obtain inclusive parton densities~\cite{Mendizabal:2023mel} and also for describing the small transverse momentum region of Drell-Yan lepton pairs in \Pp\Pp\ collisions at high energies~\cite{Bubanja:2024puv,Bubanja:2024crf,Bubanja:2023nrd}. We claim that the correlation at parton level obtained from standard MC event generators is absent, 
because soft gluons are neglected during the  initial state parton shower evolution,  while those soft gluons are effectively handled by the hadronization models; for example, in the Lund string model, gluons act as a kink in the color string.  The Parton Branching approach (\PB )~\cite{Hautmann:2017fcj,Hautmann:2017xtx} offers the possibility to study  soft gluon effects in detail. In Ref.~\cite{Mendizabal:2023mel} the effect of removing soft gluons on inclusive parton densities is shown. 


The outline of this paper is as follows. In Sec.~2 we discuss the relation of number density of gluons to multiplicity of gluons and entropy. 
In Sec.~3 we introduce Monte Carlo formulation of the DGLAP equations which is then used in the partons shower. We argue that the entropy crucially depends on the soft gluon contribution.
We then present the main results of the paper i.e. the calculation of the entropy at hadron level and parton level. We show that while the hadron level simulations describe data very well, the main contribution is due to the  hadronization mechanism.
However, when the cut on soft gluons is lowered, the dominant contribution comes from
parton density. Conclusions are presented in Sec.~4.
An appendix shows more details on soft gluon radiation in parton showers.

\section{Entanglement entropy  within the color dipole picture
}

{Before we turn to the discussion of parton and hadron production within a Monte Carlo formulation, we first recall the line of arguments by Kharzeev and Levin \cite{Kharzeev:2017qzs}, which underlies the determination of the entropy in DIS at low $x$, making use of the color dipole picture and entanglement; see also \cite{Kharzeev:2021nzh,Liu:2022hto,Hentschinski:2024gaa} for further details.}

During the interaction with the virtual photon, coherence of the proton wave function i.e. the $n$ dipole system is lost; the resulting state is no longer pure and entropy is being generated~\cite{Kharzeev:2021nzh,Kharzeev:2026jkq}. The resulting density operator is then argued to be diagonal in the dipole number basis and a determination of entanglement entropy therefore possible from the probabilities to encounter $n$ dipoles in the proton. During the recent years this proposal has been explored using a 1 dimensional reduction of the 1+2 dimensional dipole model, fixing free parameters of the model through comparing the mean number of dipoles to parton distribution functions. This allowed then for a successful description of hadronic entropy obtained for both a large fixed rapidity window \cite{Hentschinski:2021aux,Hentschinski:2022rsa}, a small  moving rapidity window  \cite{Hentschinski:2024gaa} as well as for diffractive DIS \cite{Hentschinski:2023izh}. While the studies cannot exclude with certainty that entropy is produced through a different mechanics, the ability to describe hadronic entropy through initial state dynamics has been interpreted as evidence for the close relation between hadronic entropy and entanglement.


 For a suitable choice of the basis of the observed and unobserved Hilbert spaces, called the Schmidt basis, the density matrix of the observed system $\rho$ is diagonal, and von Neuman entropy can be expressed as
\begin{equation}
\label{eq:entropy0}
S
= -\,\mathrm{Tr}\,\rho \ln \rho
= -\sum_{n} p_{n} \ln p_{n},
\end{equation}
with $ p_n \geq 0$, $\sum_n p_n=1$  eigenvalues of the reduced density matrix $\rho$.  As shown in \cite{Kharzeev:2026jkq} 
 this entropy quantifies the loss of information associated with inaccessible degrees of freedom (phases) and is directly tied to the incoherent nature of the observed system of partons. Because of that, the entropy is realized as entropy of a classical system  which in our case is the entropy of partons. 
In Ref.~\cite{Kharzeev:2017qzs} it is proposed to use the number of color dipoles probed in the proton wave function at low $x$ as a suitable realization of the  Schmidt basis. 
In DIS at low $x$, the proton wave function is characterized by a large number of quarks and gluons and a suitable approximation for this wave function within a leading logarithmic approximation is provided by the color dipole picture \cite{Mueller:1994gb}. In particular, using a large $N_c$ approximation, copious production of gluons in the low $x$ limit of DIS is within this framework described through the branching of color dipoles.  Ref.~\cite{Kharzeev:2017qzs} finally identifies the $p_n$ of Eq.~\eqref{eq:entropy0}  as the probability to encounter $n$ color dipoles in the low $x$ proton.

In \cite{Kharzeev:2017qzs} a solution for a one-dimensional reduction of the color dipole picture has been used, where all color dipoles are assumed to carry identical transverse size of the order of $~1/Q$, whereas Ref.~\cite{Liu:2022hto} presented results within a double logarithmic approximation  where  the dependence in the hard scale $Q^2$ is also taken into account.
Within the one-dimensional reduction used in \cite{Kharzeev:2017qzs}, it is then necessary to relate the $p_n$ to parton distribution functions, which allows  to  fix free parameters of the $p_n$ distribution. 
 If one relates the dipole multiplicity at certain values of $Q^2$ and $x$ to the number of gluons per unit rapidity through
\begin{equation}
\langle n(x,Q^2)\rangle=\sum_n n p_n\equiv xg(x,Q^2),
\end{equation}
(see also the discussion in \cite{Hentschinski:2022rsa}) one obtains  in the  low $x$ limit:
\begin{equation}
\label{eq:entropy}
S=\ln xg(x,Q^2) + C,
\end{equation}
where $C$ denotes a constant of order one; for the one-dimensional reduction of the dipole model one finds $C = 1$,  while the double logarithmic approximation yields $C\simeq 0.724$. In the proposal made in \cite{Kharzeev:2017qzs} it is finally suggested that it is possible to study the DIS entropy of partons through the Shannon entropy determined from the multiplicity distribution of final state charged hadrons, which have been extracted by the H1 collaboration~\cite{H1:2020zpd}. In \cite{Hentschinski:2021aux,Hentschinski:2022rsa} it is shown that a description of hadronic entropy using this approach is indeed possible (at the very least within the given uncertainties)  if one rescales  $xg(x,Q^2) \rightarrow \tfrac{2}{3} xg(x,Q^2)$, in order to take into account that only charged hadrons are measured in  experiment.

While previous phenomenological studies use sometimes slightly different methods to determine entanglement entropy \cite{Hentschinski:2021aux,Hentschinski:2022rsa,Hentschinski:2023izh,Hentschinski:2024gaa}, they all conclude that fixing free parameters of the dipole distribution from parton distribution functions allows for a successful description of the measured Shannon entropy of hadrons, at least within both experimental and theory uncertainties. Here, differences in the determination of  Eq.~\eqref{eq:entropy},  amount  essentially to a) including both the sum of gluon and quark distributions and b) to set the constant $C$ to zero or to use directly a mean number of dipoles normalized to $n(x=1, Q^2)=1$  (see also \cite{Tu:2019ouv} for a study related to proton-proton scattering).




\section{Evolution equation and parton shower and simulation of entropy}

In the following we will explore to which extend suitable results for $p_n$ can be generated from a DGLAP based parton shower within a Monte Carlo setup and whether the resulting partonic Shannon entropy can be used to describe hadronic entropy. At the moment it is not clear whether the resulting entropy can be directly interpreted as entanglement entropy. At the very least, we deal in the case of partonic entropy with entropy generated at time-scales of $~1/Q$ i.e. during the interaction with the virtual photon with $n$ parton probabilities independent of the evolution towards the hadronic final state.
To set the framework, we briefly review in the following the solution to DGLAP evolution equations with a Monte Carlo method  as well as the determination of emission and no-emission probabilities. 

\subsection{Evolution equation and parton shower\label{equations} }

The DGLAP evolution equation for the momentum-weighted densities  $xf_a(x,\mu^2)$ 
of parton $a$ with momentum fraction $x$ at the scale $\mu$ reads
\begin{equation}
\label{EvolEq}
 \mu^2 \frac{{\partial }{x f}_a(x,\mu^2)}{{\partial } \mu^2}   =  
 \sum_b
\int_x^{1} dz \; {P}_{ab} \left(\as(\mu^{2}),z\right)  \; \frac{x}{z}{f}_b\left({\frac{x}{z}},
\mu^{2}\right)   \; .
\end{equation}
The regularized DGLAP splitting functions $P_{ab}$ describe the splitting of partons $b$ into  $a$ and are summarized for NLO and NNLO in Ref~\cite{\nnloSplit}. 
The plus-prescription in the regularized splitting functions can be replaced by a Sudakov form factor $\Delta^S_a(z_M, \mu^2)$, as applied in 
the \PB -approach~\cite{Hautmann:2017fcj,Hautmann:2017xtx}, and the evolution equation can be rewritten  as
\begin{equation}
\label{EvolEqSudakov}
  {x f}_a(x,\mu^2)  =  \Delta^S_a (  \mu^2  ) \  {x f}_a(x,\mu^2_0)  
+ \sum_b
\int^{\mu^2}_{\mu^2_0} 
{{d q^2 } 
\over q^2 } 
{
{\Delta^S_a (  \mu^2  )} 
 \over 
{\Delta^S_a( q^2
 ) }
}
\int_x^{\zM} {dz} \;
 \frac{\as}{2\pi} \hat{P}_{ab} (z) \frac{x}{z}
\;{f}_b\left({\frac{x}{z}},
q^2\right)  \; ,
\end{equation}
where the Sudakov form factor, 
\begin{equation}
\label{sud-def}
  \Delta_a^S ( \mu^2 ) \equiv \Delta_a ( \mu^2 , \mu^2_0 ) = 
\exp \left(  -  \sum_b  
\int^{\mu^2}_{\mu^2_0} 
{{d { q}^{ 2} } 
\over {q}^{2} } 
 \int_0^{\zM} dz  \frac{\as}{2\pi}
\ z  P_{ba}^{(R)}\left( z \right) 
\right) 
  \;\; ,   
\end{equation}
sums up unresolved real and virtual emissions.
Both in the hard emission part of Eq.~\eqref{sud-def} and in the Sudakov form factor, the upper limit $\zM$ of the $z$-integral is essential for the following discussion: soft gluons live in the region of  $z\to1$. This can be directly seen in an angular ordering approach, where the transverse momentum of the emitted parton is given by  $\qt  =  (1 - z ) q $.

In a Monte Carlo event generator, like \pythia \cite{Sjostrand:2014zea,Bierlich:2022pfr}, the initial-state parton shower  starts, for efficiency reasons,  from the hard scattering and evolves backwards to the hadronic scale. For this backward evolution, a Sudakov form factor $\Delta_{bw}$ is used, which is different from the one in the evolution of parton densities, as the backward evolution is guided by the parton densities (see for example the discussion in Ref.~\cite{Ellis:1991qj}). The  Sudakov form factor $\Delta_{bw}$, which includes the parton densities, reads
\begin{equation}
\Delta_{bw} (z, \mu^2, \mu^2_{i-1}) = \exp \left( - \sum_b \int_{\mu_{i-1}^2}^{\mu^2}  \frac{d q^{\prime\,2}}{q^{\prime\,2}}\int_x^{\zM} dz P^{(R)}_{ab} (\as(z,q'),z) \frac{x' f_b(x',q')}{xf_a(x,q')} \right)\; .
\label{Suda}
\end{equation}
In both cases, the form factor gives the probability for no emission between the scales $ \mu$ and $\mu_{i-1}$ in the chain.

The probability for any emission is related to the probability for no emission by unitarity. 
With this relation, the probability for emissions during an initial state evolution is directly related to the parton density, as one can see  from the formulation of the parton density in terms of a Sudakov form factor Eq.~(\ref{EvolEqSudakov}). Both the no-emission probability and the emission probability depends  therefore on $\zM$, i.e. the treatment of soft gluons. Thus, any deviation from $\zM \to 1$ will lead to a different number of soft gluons and therefore to a different number of partons produced in the initial state cascade. Therefore,
from the self consistency of the evolution equation, one argues that also the unresolved gluons and virtual emissions play a role in the entropy production in order to preserve  unitarity of the evolution.
We would like to point out that for an inclusive formulation of DGLAP evolution this $z\rightarrow 1$ singularity of the (real part of the) splitting function  is usually regulated through a plus-prescription. 

To finally briefly illustrate  the effect of finite $z_M$, we determine the gluon distribution with no emissions, $xg_{ne}(x,\mu^2)$, i.e. the starting distribution  together with resummed unresolved and virtual corrections  only,  provided by the first term of the RHS of Eq. \eqref{EvolEqSudakov}. To allow for an analytic evaluation, we ignore here powers of $(1-\zM)$ 
One finds
\begin{align}
xg_{ne}(x, \mu^2)
&= \Delta(\mu^2)\, xg(x\mu_0^2) \notag \\ 
&=xg(x,\mu_0^2) \exp\left\{ - \int_{\mu_0^2}^{\mu^2} 
    \frac{dq^{2}}{q^{2}} 
    \int_0^{z_M} dz \; z \, 
    \frac{\as}{2\pi}\, \left[
    P_{gg}^{(R)} +  P_{gq}^{(R)} \right]\!\left( z \right) \right\}
      \notag \\
&= 
xg(x,\mu_0^2) \exp\left\{\ln\!\frac{\mu^2}{\mu_0^2} \frac{\as C_A}{\pi}\,\left[
   \frac{11}{12}
   +
   \!\ln(1-z_M)\right]  -\ln\!\frac{\mu^2}{\mu_0^2}\frac{\as C_F}{\pi} \frac{2}{3}\right\}.
\end{align}
Since $\ln(1-z_M)<0$, the term depending on $z_M$ gives a negative contribution, and  in the limit  $z_M \to 1$, this contribution completely suppresses the gluon distribution without emission. Through unitarty and through the need to obtain identical global parton distributions for any value of $\zM$, the limit $z_M \to 1$ therefore naturally increases the emission probability and as a consequence the number of resolved partons. At the level of entropy, which is directly related to the number of resolved real emissions, this translates into an increase in entropy for $\zM$ approaching one.

\subsection{Simulations for particle and parton multiplicities}

\begin{figure} [t]
\centering
\includegraphics[width=0.75\linewidth]{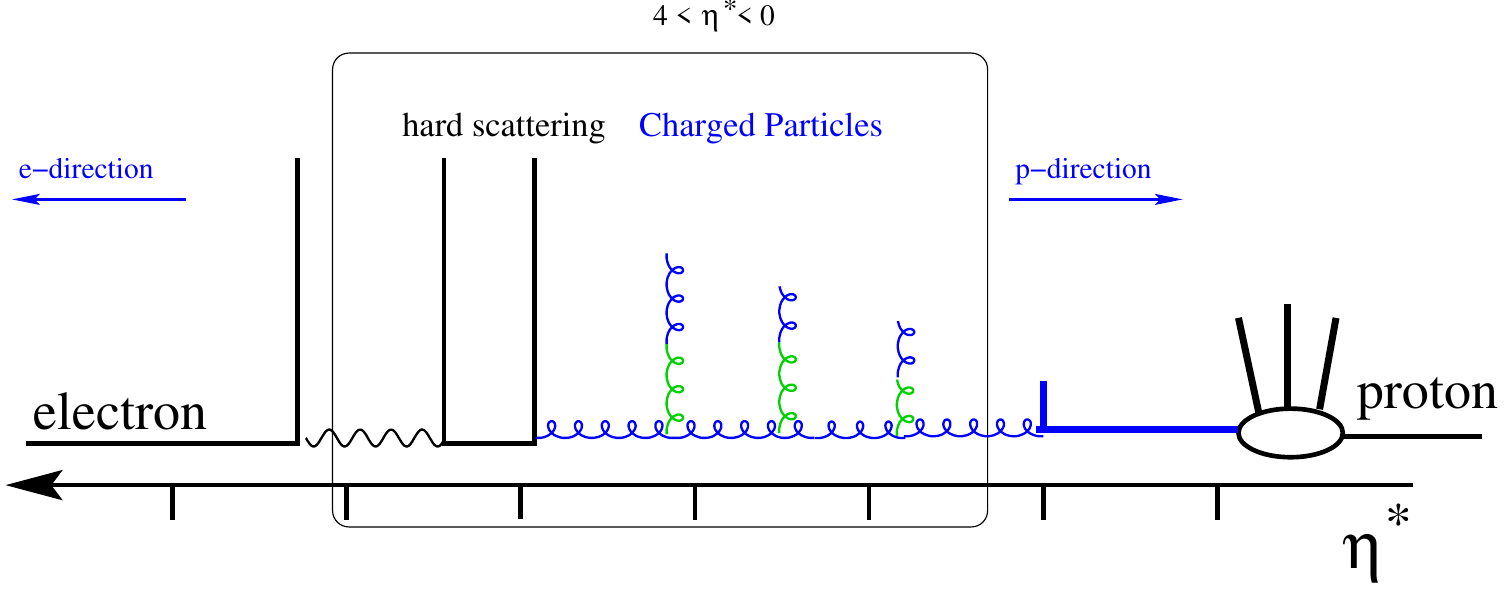}
\caption{ The different pseudorapidity regions in a DIS process. Indicated is also the range in pseudorapidity in the hadronic center-of-mass frame. Soft gluons are indicated in "green".}
\label{fig:DIS-diagram}
\end{figure}

The H1 collaboration~\cite{H1:2020zpd} has measured charged particle multiplicities and calculated the hadron entropy $S_{\rm hadron}$ in different regions of  the pseudorapidity of the charged particles for $\pt > 0.15$~\GeV\ in various bins of $x$ and  $Q^2$. The entropy eq.(\ref{eq:entropy0}) is calculated from the mean multitplicity in bins od $x$ and $Q^2$, either on hadron or on parton level. An illustration of the rapidity region and the hard scattering process in DIS is shown in Fig.~\ref{fig:DIS-diagram}.

We use a full NLO simulation in DIS obtained with \powheg~\cite{Banfi:2023aa} supplemented with parton shower and hadronization. The simulated events are analyzed using Rivet~\cite{Bierlich:2019rhm}. The parton shower is provided within the \pythia\ framework. For the presented study we use the improved parton shower \pythia -\pythiaPB~\cite{Jung:2025mtd}, which  ensures, that the parton shower is fully consistent with the DGLAP evolved collinear PDF (which in general is not the case in standard parton showers). For completeness we also  show results obtained with \rapgap~\cite{Jung:1993gf,RAPGAP33}, which has been used in the H1 analysis, and we compare results with a Monte Carlo simulation based on the CCFM small $x$ evolution~\cite{\ccfm} equation obtained with CASCADE~\cite{Jung:2001hx,cascadeweb,Jung:2010si,Baranov:2021uol}

\subsubsection{Entropy at hadron level}
\begin{figure} [t]
\centering
\includegraphics[width=0.45\linewidth]{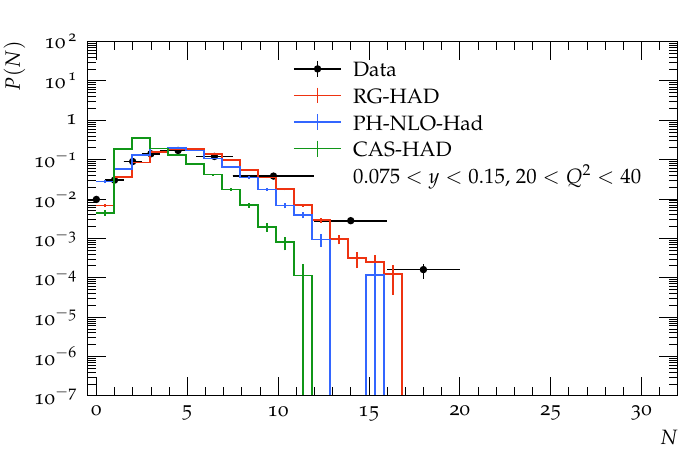}
\includegraphics[width=0.45\linewidth]{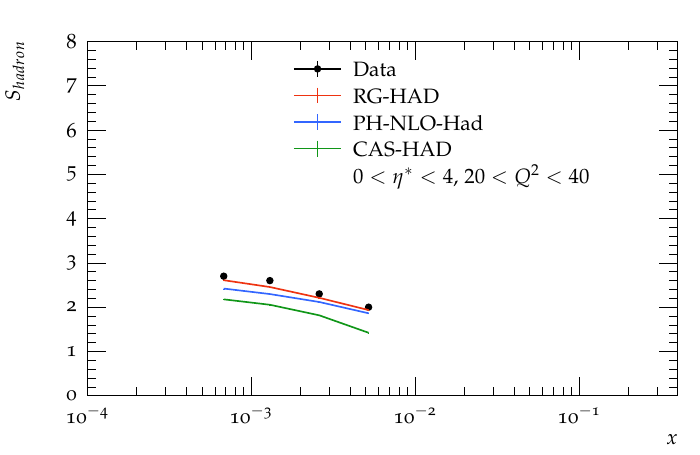}  
\caption {\small Left: Charged particle multiplicity at $20 < Q^2 <40$~\GeV$^2$. Right: Entropy $S_{\rm hadron}$ as a function of $x$. Shown are the predictions obtained with \protect\rapgap , \protect\powheg-\pythiaPB\ and \cascade , the measurement is from H1~\protect\cite{H1:2020zpd}.}
\label{fig:RapgapPowheg}
\end{figure}

The predictions at hadron level are obtained from full simulations applying the same cuts as in the experimental analysis. 
In Fig.~\ref{fig:RapgapPowheg}(left)  we show the distribution of charged particles  (for $\pt > 0.15$~\GeV ) and in Fig.~\ref{fig:RapgapPowheg}(right) we show entropy of hadrons, calculated according to Eq.~\eqref{eq:entropy0},  evaluated at  $20<Q^2<40$~GeV$^2$. The complete set of $Q^2$ bins is shown in appendix~\ref{ChargedPartcileAddPlots}. The predictions are obtained using  \powheg\ DIS at NLO~\cite{Banfi:2023aa}  supplemented with  \pythia -\pythiaPB~\cite{Jung:2025mtd},  \rapgap~\cite{Jung:1993gf,RAPGAP33} (as in the H1 publication~\cite{H1:2020zpd})  and   CASCADE~\cite{Jung:2001hx,cascadeweb,Jung:2010si,Baranov:2021uol} based on the CCFM small $x$ evolution~\cite{\ccfm}, which describes DIS with only gluons in addition to valence quarks.


All predictions give a reasonably good description of the measured charged particle distributions, although very different theoretical models, i.e. the collinear factorization based models \rapgap~(LO) and \protect\powheg-\pythiaPB~(NLO)  and a  CCFM/\kt -factorization model have been used.

\begin{figure} [t]
\centering
\includegraphics[width=0.45\linewidth]{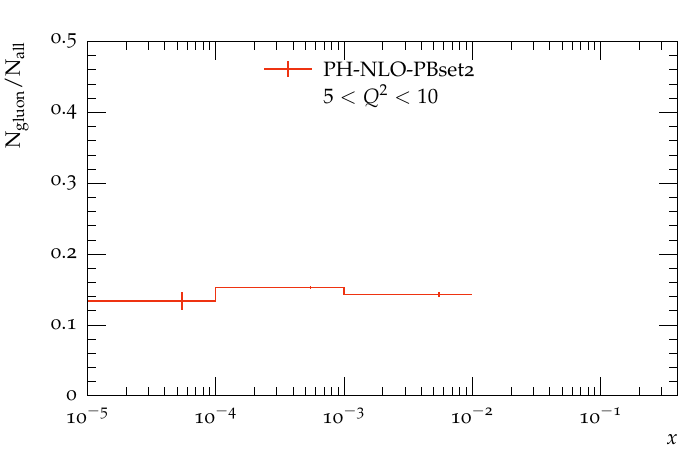}
\includegraphics[width=0.45\linewidth]{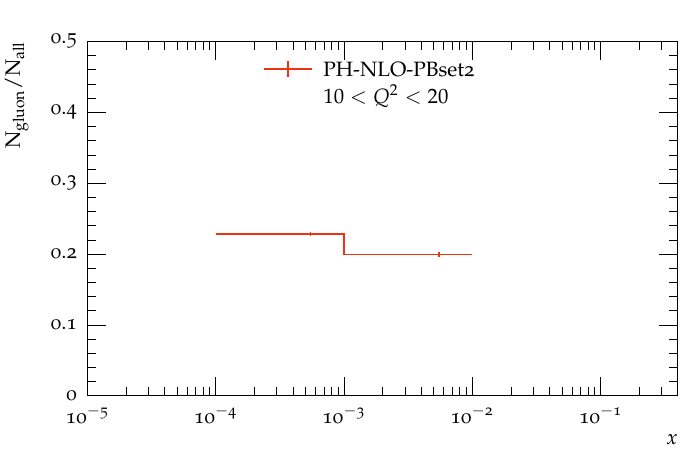}  
\caption {\small Fraction of gluon induced processes in the kinematic region of the H1 measurement~\protect\cite{H1:2020zpd} for $5 < Q^2 <10 $, $10 < Q^2 <20$ \GeV$^2$. Shown are the predictions obtained with \protect\powheg-\pythiaPB . }
\label{fig:PowhegGluon}
\end{figure}

In Fig.~\ref{fig:PowhegGluon} we show the fraction of gluons which induce the hard processes in the kinematic region of the H1 measurements as obtained from the \powheg\ simulation.  Clearly, as this is a DIS process where the photon  couples directly to quarks, quark induced processes are dominant even at NLO, where hard gluons start to contribute. 
The situation is different in calculations based on \kt -factorization and CCFM evolution, where hard gluons are relevant already at LO. 
However, independent of the hard process, as we will show in the next section, soft gluon emissions (from both quark and gluon initial states) are  important for soft particle correlations rather than the initiating parton.

\subsubsection{Entropy at parton level}
While a proper description of the experimental measurements can only be achieved at hadron level, we study next the entropy on parton level, in order to investigate the dependence of the entropy as a function of $x$.
We want to answer the question, whether  the rise of entropy with decreasing $x$ is an effect of hadronization, or whether it can be traced back to soft gluons at parton level.

While for the experimental analysis of charged particle multiplicities, charged particles with $\pt > 0.150$~\GeV\ are selected, we count for the parton level multiplicities all partons within a certain rapidity range (as in the H1 analysis), but without a minimum \pt\ requirement. We attempt to study the influence of restrictions in the parton shower such as the cut on \qt\ (the limit on \zM\  on parton level which regulates the amount of soft gluon emissions in angular ordering) on the parton multiplicities and $S_{\rm parton}$. It is obvious, that a restriction of soft gluon emissions leads to a reduction of the parton multiplicities. 

\begin{figure} [t]
\centering
\includegraphics[width=0.47\linewidth]{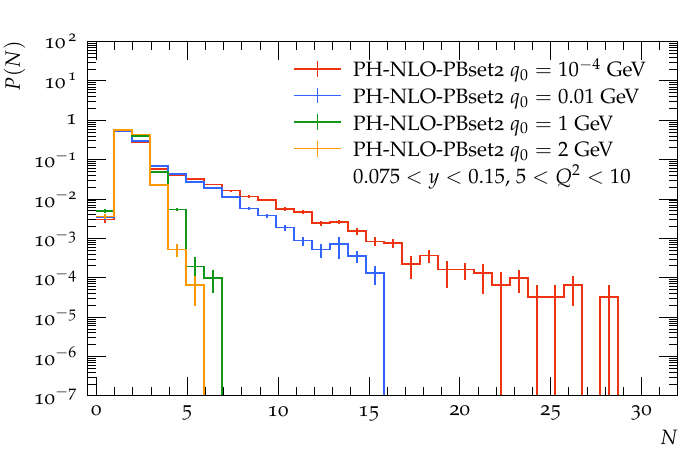}
\includegraphics[width=0.47\linewidth]{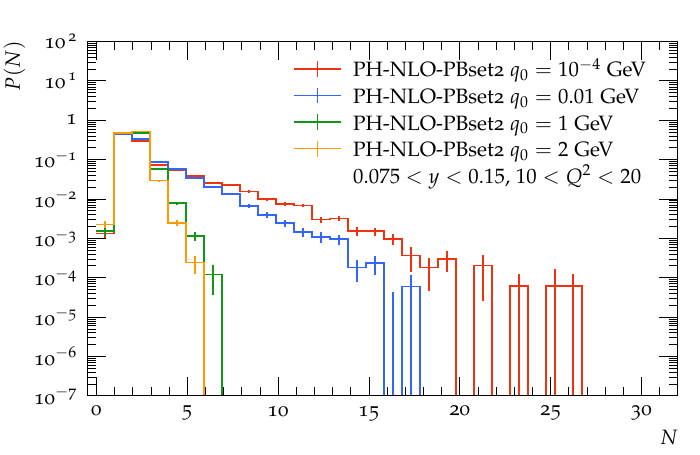}
\includegraphics[width=0.47\linewidth]{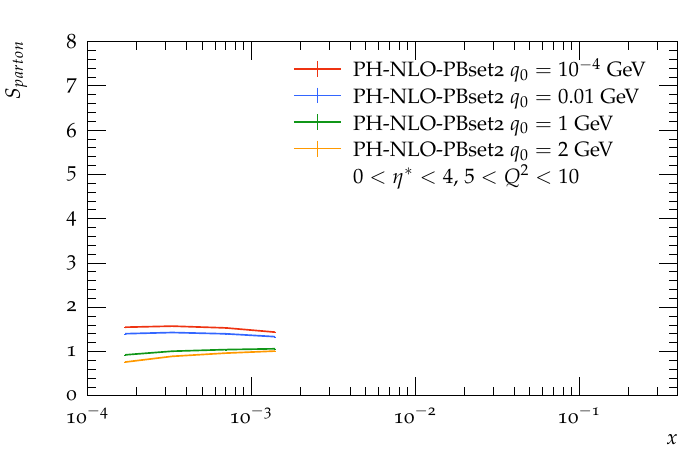}  
\includegraphics[width=0.47\linewidth]{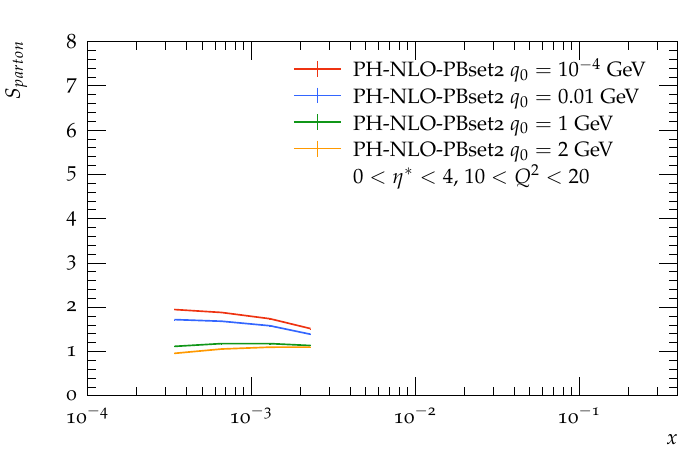}  
\caption {\small Upper row: Partonic multiplicity at  $5 < Q^2 <10 $ and $10 < Q^2 <20 $  \GeV$^2$ .  Lower row: Entropy $S_{\rm parton}$ as a function of $x$. Shown are the predictions obtained with \protect\powheg-\pythiaPB\ for $q_0=10^{-4}$, $0.01$, $1$ and $2$ \GeV .}
\label{fig:Powheg_q0}
\end{figure}

In Fig.~\ref{fig:Powheg_q0} predictions obtained with \powheg-\pythiaPB\ are shown for different $q_0$ which limits the transverse momentum of partons during the initial state shower: $\zM = 1- q_0/q$.
In \pythia\ soft gluon emission in the initial state parton shower is restricted due to cuts on \zM\ as well as the ordering condition (\pt -ordering in \pythia\ vrs angular ordering in \powheg-\pythiaPB). A more detailed discussion on the restriction of soft gluon emissions can be found in appendix~\ref{SoftGluonsInPS}. 
We conclude, that  a significant \qt -cut reduces the  correlations in  $S_{\rm parton}$ significantly.
In collinear calculations of DIS (even at NLO as in  \powheg-\pythiaPB ), the contribution of initial gluons is important, but the contribution of quarks cannot be neglected. The situation is different in \cascade , where no sea-quarks are involved, and the gluon density (the CCFM gluon density) plays the dominant role. 

\begin{figure} [t]
\centering
\includegraphics[width=0.47\linewidth]{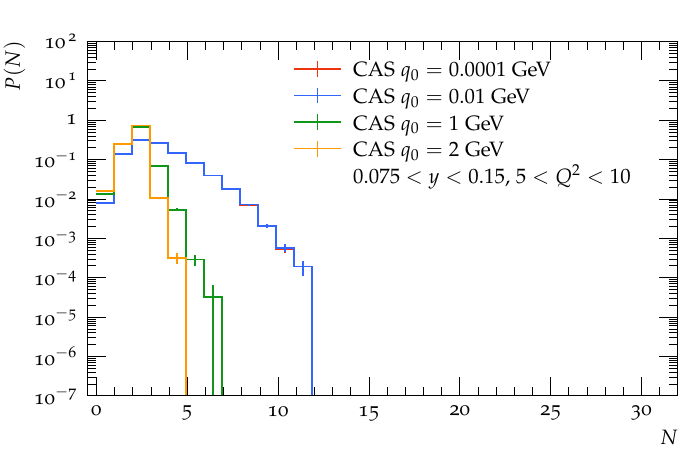}
\includegraphics[width=0.47\linewidth]{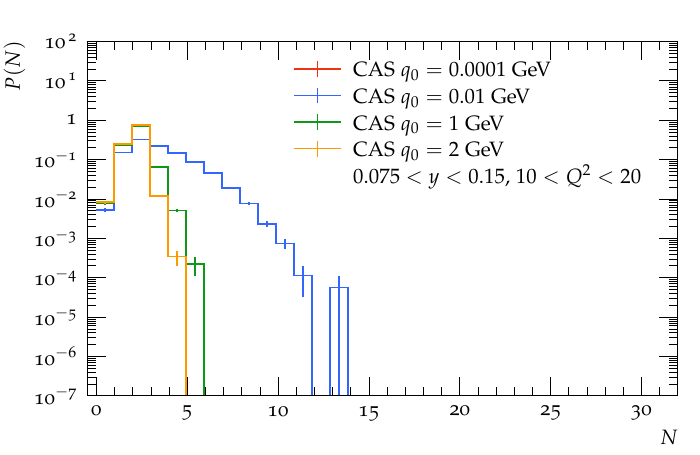}
\includegraphics[width=0.47\linewidth]{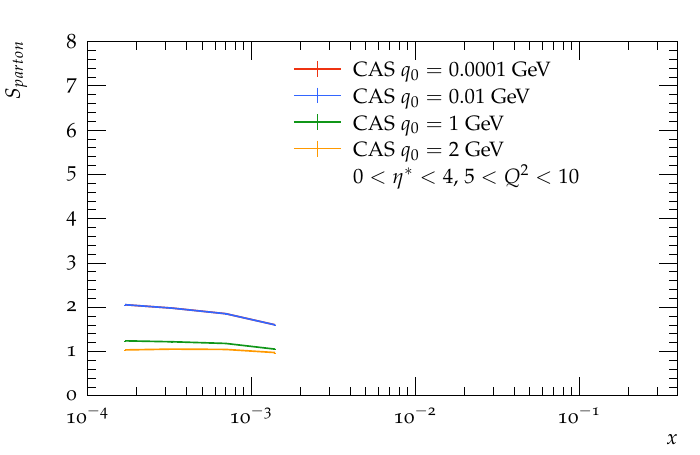}  
\includegraphics[width=0.47\linewidth]{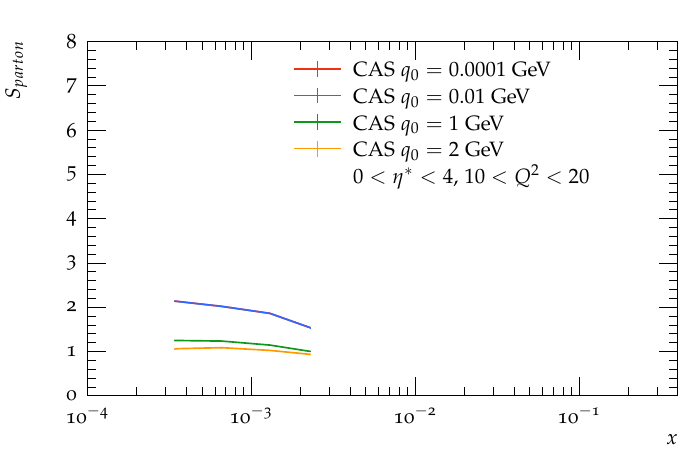}  
\caption {\small Upper row: Partonic multiplicity at  $5 < Q^2 <10 $ and $10 < Q^2 <20 $ \GeV$^2$ . Lower row: Entropy $S_{\rm parton}$ as a function of $x$. Shown are the predictions obtained with \protect\cascade\ for $q_0=10^{-4}$, $0.01$, $1$ and $2$ \GeV . The prediction for $q_0=10^{-4}$ \GeV\ is almost identical to the one with $q_0=0.01$ \GeV . }
\label{fig:CASCADE_q0}
\end{figure}

In Fig.~\ref{fig:CASCADE_q0} we show the parton multiplicity distributions and $S_{\rm parton}$ as a function of the soft-gluon cutoff parameter $q_0$ obtained with \cascade . Essentially the same dependence, as observed with  \powheg-\pythiaPB , is obtained: less restrictions on soft gluon emissions (with a lower $q_0$ parameter) leads to a larger multiplicity of partons.

While at parton level, the parton multiplicity rises with smaller $q_0$, stability is reached at hadron level, since very soft gluons give only a small kink to the color string in string fragmentation. This stability has been already reported in a study on the role of soft gluons in Drell-Yan production~\cite{Mendizabal:2023mel}.

We conclude that the proximity of $S_{\rm parton}$ and measured $S_{\rm hadron}$ originate from soft gluon emissions, which are crucial not only for parton densities but also for parton and charged-particle multiplicities.

A natural question to ask is whether a description in terms of partons is meaningful, if transverse momenta take values as small as $q_0 = 10 ^{-4}$~GeV. Indeed, partons at very small momenta need special attention, since they cannot be treated in perturbative QCD alone, and require at the very least resummation. The parton shower approach we are applying treats theses partons in a way consistent with collinear and transverse momentum dependent parton densities, including an extension of $\alpha_s$ into the non-perturbative region (as described in Refs.~\cite{Hautmann:2017fcj,Hautmann:2017xtx}). 
We therefore believe that partonic entropy, as determined through the parton shower, provides at the very least a first approximation to entanglement entropy, which is generated during the interaction of the virtual photon with the proton i.e. at time scales of the order of  $\sim 1/Q$. The resulting partonic system is  characterized by high parton virtualities and  lies therefore within  the realm of perturbative QCD. Our result is therefore based on the assumption that the entropy of this partonic system is well approximated through the system created by the parton shower which itself provides an exclusive realization of DGLAP evolution.  We further stress that full agreement of the parton shower with (inclusive) DGLAP evolution and therefore consistency with inclusive (collinear) parton densities is only achieved in the limit $q_0 \to 0$. 

From a more general perspective, the following interpretation is furthermore possible: During the interaction of the proton with the virtual photon, the pure quantum state of the proton turns into a mixed state since coherence in the proton wave function is lost; the mixed state is then described by a certain density operator $\rho$. If degrees of freedom in the proton are entangled, entanglement entropy is generated and determined through $S = \text{tr} \rho \ln \rho$. While the density operator transforms under unitary transformations $U$, such as quantum mechanical time evolution, as $\rho \to U^\dagger \rho U$, entanglement entropy itself is invariant under unitary transformations. Entropy generated during the interaction with the virtual photon is according to this argument unmodified by evolution towards the final state. We currently do not know whether this property does hold in this form in DIS. In Monte Carlo calculations this property seems to be restored at the parton level, once soft gluons are taken into account. Even though partonic and hadronic results do not completely agree and hadronization is required for a precise description of not only multiplicities but also hadronic entropy, partonic entropy reproduces correctly the rise with $1/x$ and approaches in general the hadronic result, once soft partons are included.



In order to have a more complete understanding of the interplay of hard and soft emissions, it would be interesting to consider entropy and multiplicity distributions at higher center of mass energies, such as at a future high energy $ep$-collider  or  within $pp$ collision in forward processes at the Large Hadron Collider \cite{ALICE:2023fov}. In such a scenario one expects multiple hard emissions in the low $x$ limit and it would be interesting to study their interplay with soft emissions in the generation of entropy.

\section{Conclusion and outlook\label{Conclusion}}

We have studied charged particle and parton multiplicities in DIS and calculated the  Shannon hadron $S_{\rm hadron}$ and $S_{\rm parton}$  entropies. The aim was to understand whether one can obtain a rising entropy of partons as simulated in Monte Carlo event generators. 
We found that at hadron level the distributions can be described essentially with Monte Carlo simulations using different parton showers and hadronization. 

However, at parton level, the parton multiplicity  very strongly depends on the simulation of soft gluon emissions. While Monte Carlo event generators like \rapgap\ and \pythia\, with standard settings, do not pay attention to soft gluon emissions, and remove most of it, there is little dependence at parton level observed.

With the newly implemented parton shower \pythiaPB\ into the \pythia\ event generator, which puts emphasis on soft gluon emissions, we were able to study in detail their contribution and found, that they contribute mainly to the parton multiplicities.
A very similar behaviour is observed using the CCFM Monte Carlo generator \cascade\ which relies essentially on unintegrated gluon densities. 

This study clarifies the open question on the origin of particle multiplicities and $S_{\rm hadron}$ and traces it back to the contribution of soft gluons. It also poses a question on whether current application of hadronization effects and fragmentation function needs to be re-investigated to allow for more contribution from initial states. For the future it would be very interesting to study in more detail the role of the hadronization model and their interplay with entropy generation.

\vskip 0.5 cm 
\begin{tolerant}{8000}
\noindent 
{\bf Acknowledgments} 
We are grateful for many discussions with A.~Bagdatova, S.~Baranov,  A.~Kotikov, A.~Lipatov, M.~Malyshev, G.~Lykasov and the other participants  of the WeeklyOfflineMeeting during the past years. Furthermore we would like to thank M.~Praszalowicz, C.~Marquet and L.~Motyka for stimulating discussions.
KK acknowledges hospitality of the QCD group at École Polytechnique, Institut Polytechnique de Paris, where most of this work was completed. The work of KK was supported by NCN grant No. 2019/33/B/ST2/02588
and SSHN French fellowship for the year 2025.
\end{tolerant}

\section*{Appendix}
\section{Charged particle multiplicities and $S_{\rm hadron}$ for different $Q^2$ regions \label{ChargedPartcileAddPlots}}
\begin{figure} [htb]
\centering
\includegraphics[width=0.32\linewidth]{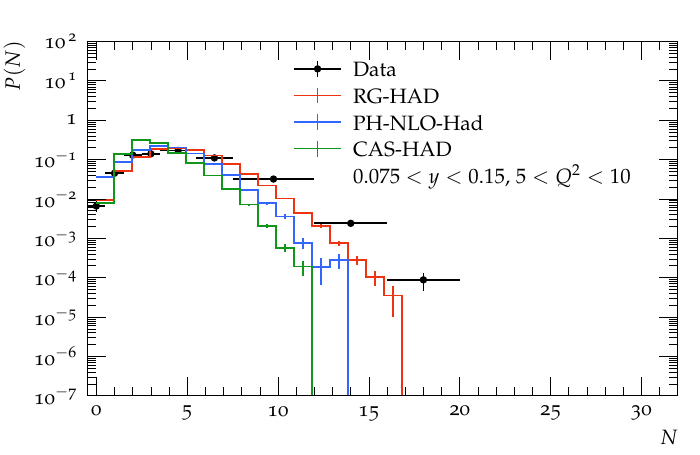}
\includegraphics[width=0.32\linewidth]{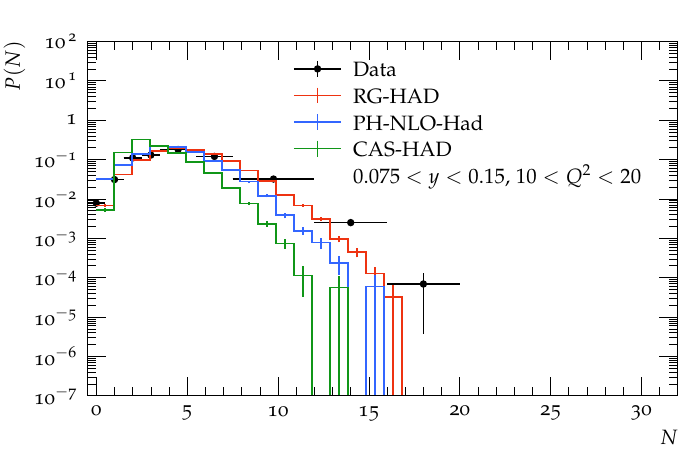}
\includegraphics[width=0.32\linewidth]{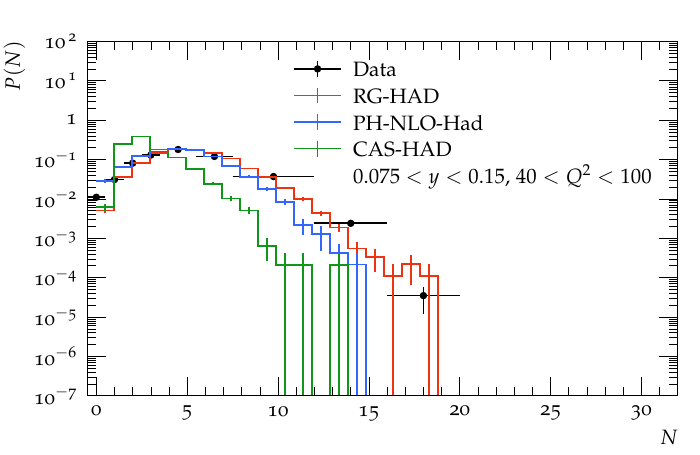}
\includegraphics[width=0.32\linewidth]{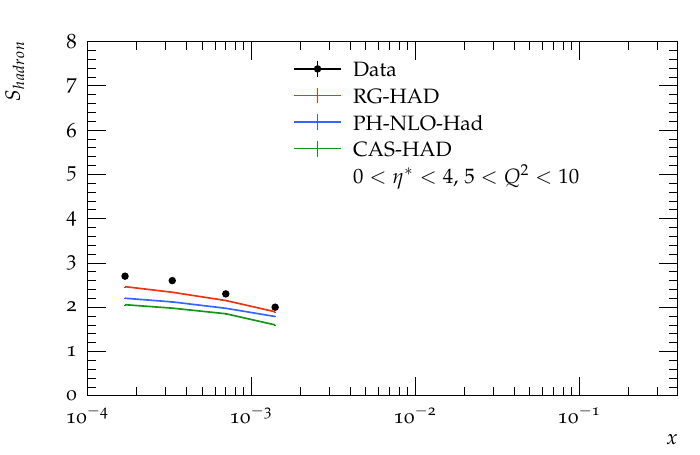}  
\includegraphics[width=0.32\linewidth]{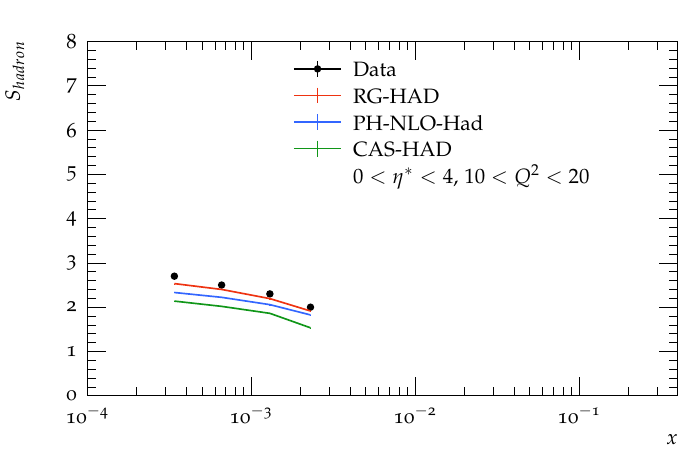}  
\includegraphics[width=0.32\linewidth]{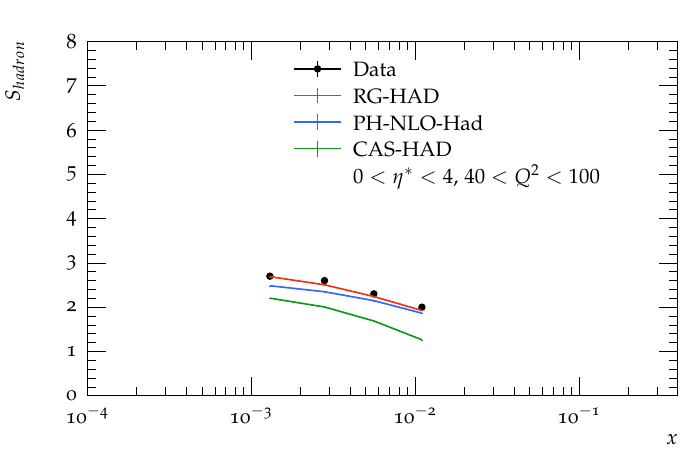}  
\caption {\small Upper row: Charged particle multiplicity at $5 < Q^2 <10 $, $10 < Q^2 <20 $ and $40 < Q^2 <100 $ \GeV$^2$. Lower row: Entropy $S_{\rm hadron}$ as a function of $x$. Shown are the predictions obtained with \protect\rapgap , \protect\powheg-\pythiaPB\ and \cascade .}
\label{fig:RapgapPowheg_add}
\end{figure}

In Fig.~\ref{fig:RapgapPowheg_add} the charged particle multiplicities and $S_{\rm hadron}$ for all four $Q^2$ regions are shown.

\begin{figure} [t]
\centering
\includegraphics[width=0.47\linewidth]{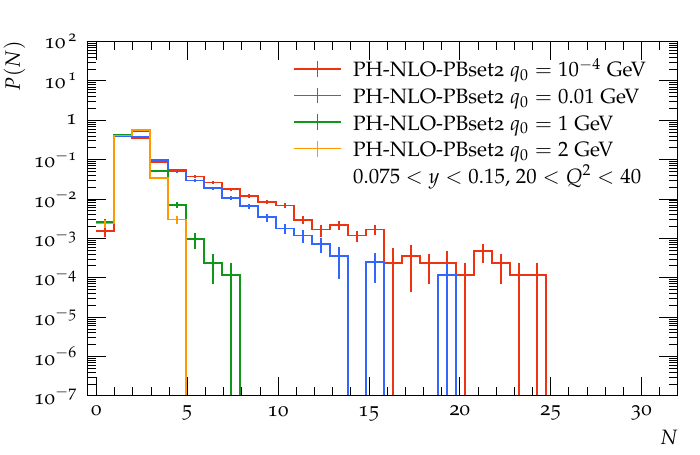}
\includegraphics[width=0.47\linewidth]{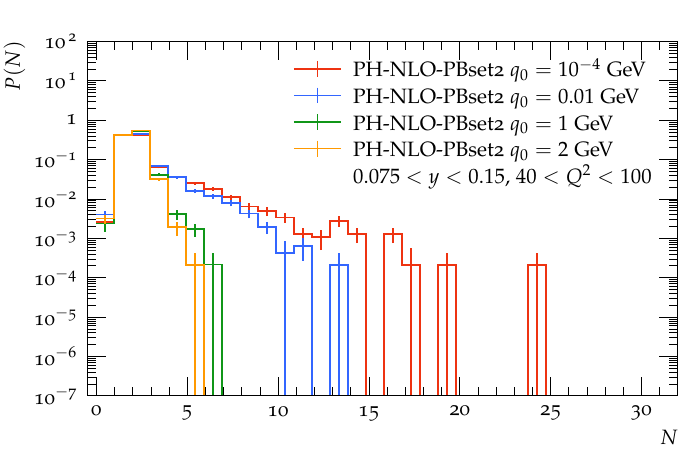}
\includegraphics[width=0.47\linewidth]{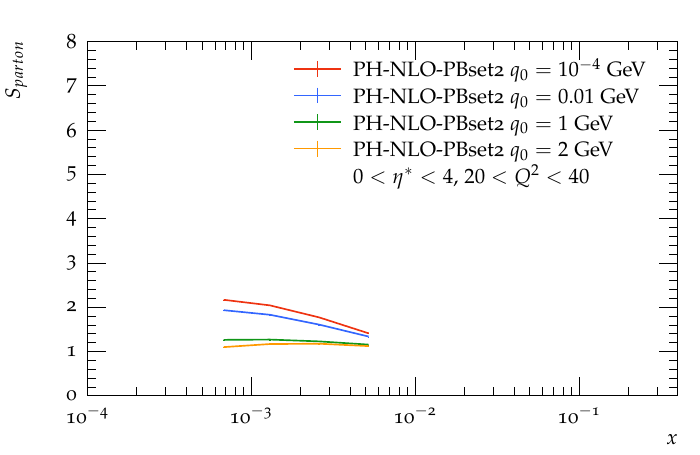}  
\includegraphics[width=0.47\linewidth]{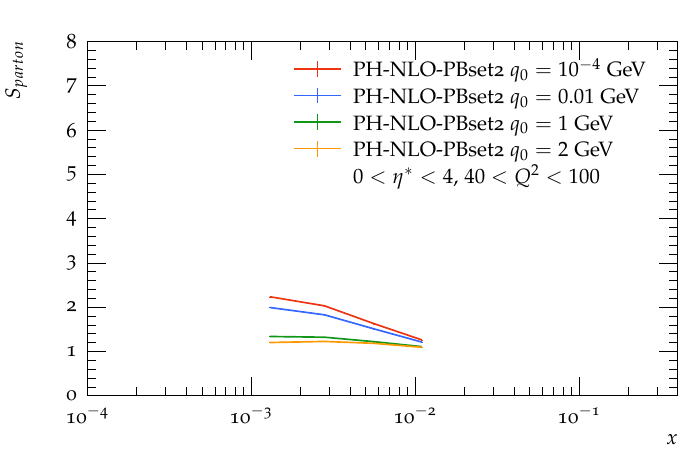}  
\caption {\small Upper row: Partonic multiplicity at  $20 < Q^2 <40 $ and  $40 < Q^2 <100 $ \GeV$^2$ .  Lower row: Entropy $S_{\rm parton}$ as a function of $x$. Shown are the predictions obtained with \protect\powheg-\pythiaPB\ for $q_0=10^{-4}$, $0.01$, $1$ and $2$ \GeV .}
\label{fig:Powheg_q0_1}
\end{figure}

In Fig.~\ref{fig:Powheg_q0_1} predictions of the partonic multiplicities and $S_{\rm parton}$  obtained from  \powheg-\pythiaPB\ are shown for different values of $q_0$. 

\begin{figure} [t]
\centering
\includegraphics[width=0.47\linewidth]{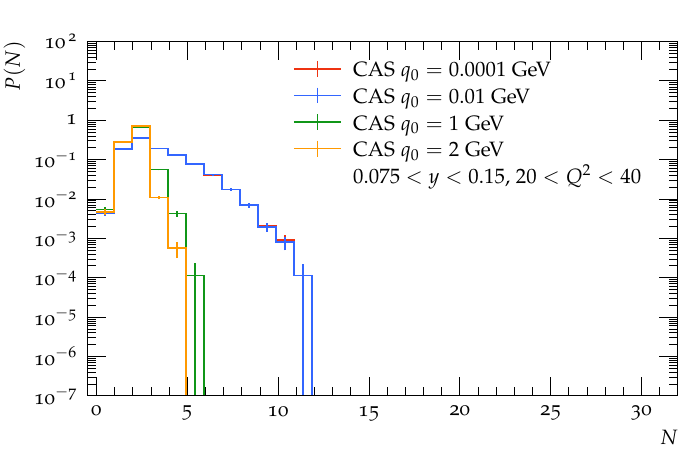}
\includegraphics[width=0.47\linewidth]{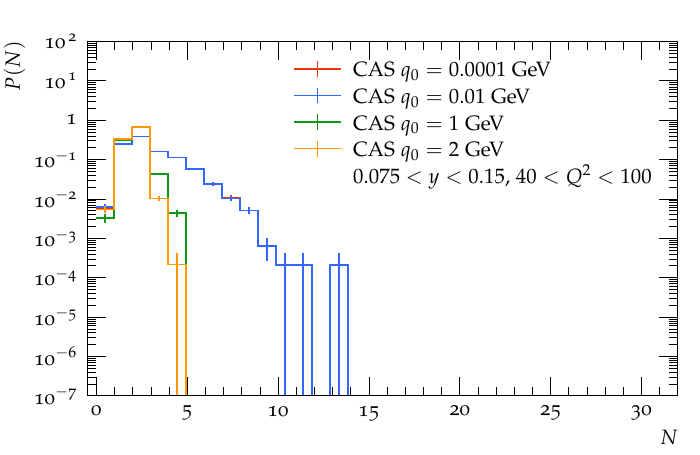}
\includegraphics[width=0.47\linewidth]{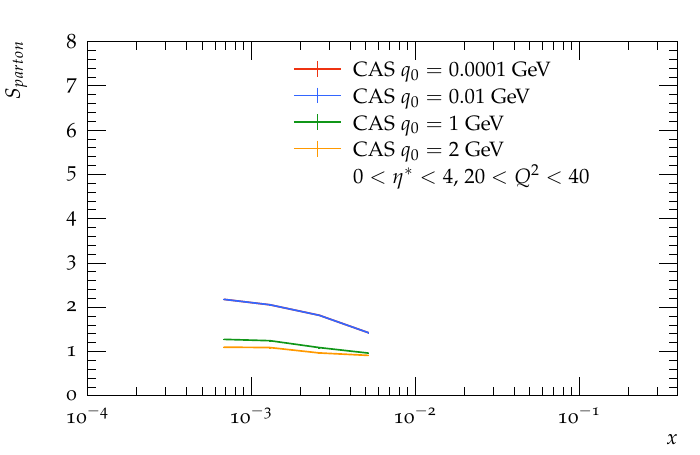}  
\includegraphics[width=0.47\linewidth]{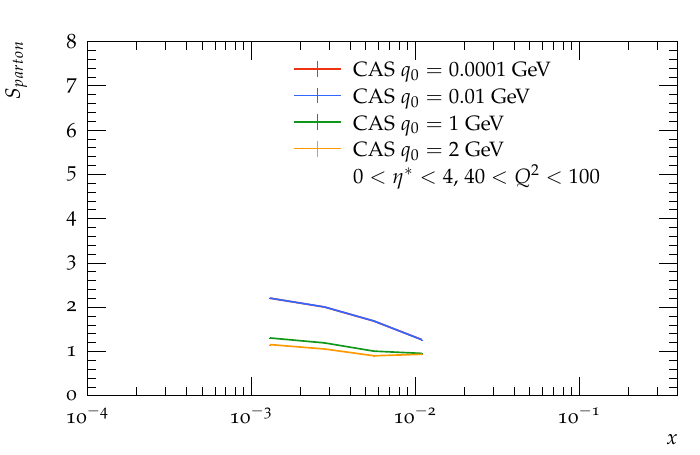}  
\caption {\small Upper row: Partonic multiplicity at   $20 < Q^2 <40 $ and $40 < Q^2 <100 $ \GeV$^2$ . Lower row: Entropy $S_{\rm parton}$ as a function of $x$. Shown are the predictions obtained with \protect\cascade\ for $q_0=10^{-4}$, $0.01$, $1$ and $2$ \GeV . The prediction for $q_0=10^{-4}$ \GeV is almost identical to the one with $q_0=0.01$ \GeV .}
\label{fig:CASCADE_q0_1}
\end{figure}

In FIg.~\ref{fig:CASCADE_q0_1} predictions   of the partonic multiplicities and $S_{\rm parton}$ obtained  with  \cascade\ are shown.

\section{Soft gluon emissions in initial state parton shower \label{SoftGluonsInPS}}
The role of soft gluons can be most easily studied in $\Pp\Pp \to \PZ\  +X $ production at large $\sqrt{s}$. We study in detail the initial state parton shower from \pythia (using tune CUET) as well as \pythia -\pythiaPB\ using \PZ -production at $\sqrt{s}=13$~\TeV . Hadronization as well as multiparton emission is turned off. The collinear parton densities are \PBset~Set2.

\begin{figure} [t]
\centering
\includegraphics[width=0.47\linewidth]{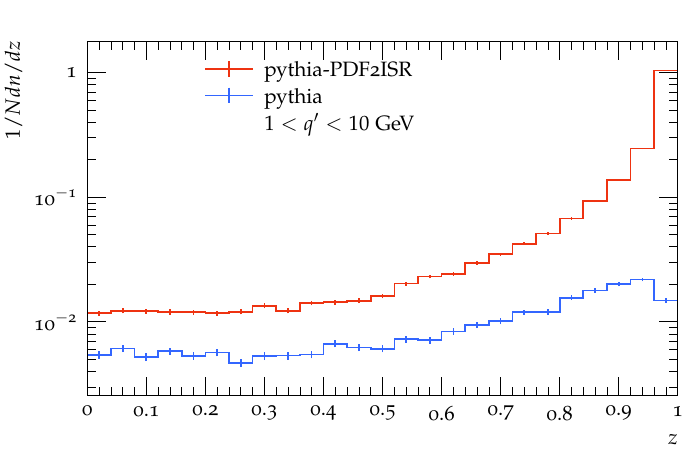}
\includegraphics[width=0.47\linewidth]{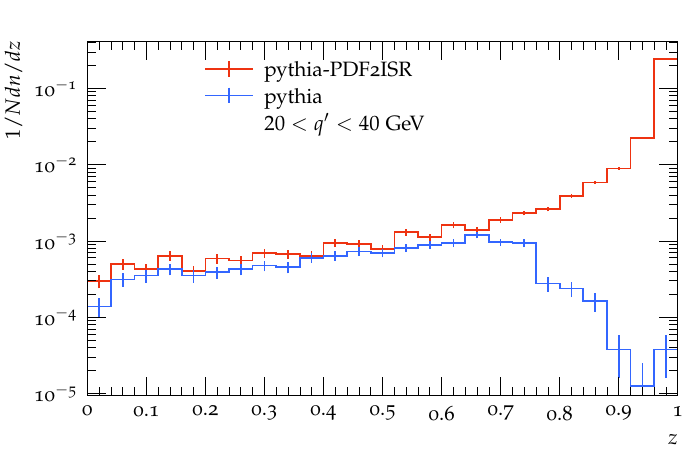}
\includegraphics[width=0.47\linewidth]{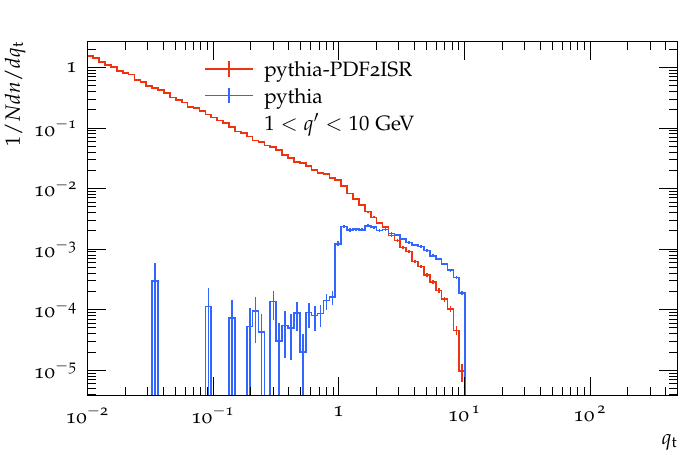}
\includegraphics[width=0.47\linewidth]{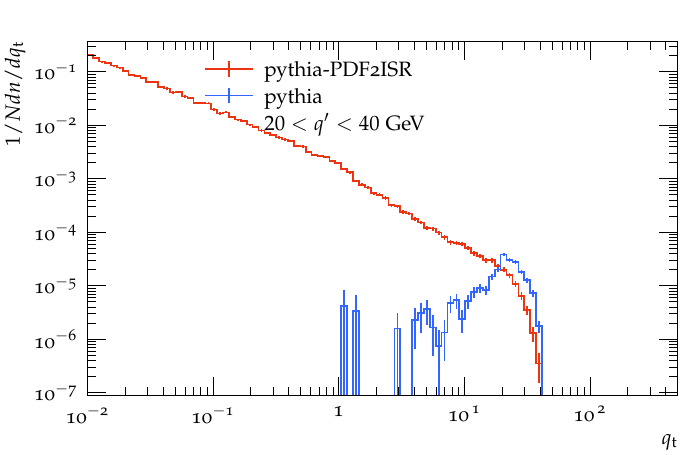}
\caption {\small Distributions obtained from \pythia-\pythiaPB\  and  \pythia\  for $1 < q' < 10 $ and $ 20 <q' < 40 $ \GeV:
upper row: Distribution in $z$, lower row: distribution in \qt .}
\label{fig:ISR_zdistribution}
\end{figure}

In Fig.~\ref{fig:ISR_zdistribution} (upper row) we show the distribution of $z$, as obtained during the initial state shower, at  different scales $q'$ (see Eq.\eqref{Suda}). While the prediction from \pythia -\pythiaPB\ shows a $z$-distribution rising towards $z \to 1$, as expected from the quark splitting function $P_{qq} \sim \frac{1}{1-z}$, the prediction from \pythia\ shows a clear deficit at $z \to 1$ for larger $q'$, only at low $q'$ the full range of $z$ is approximately covered.

In Fig.~\ref{fig:ISR_zdistribution} (lower row) we show the distribution in \qt , the transverse momentum of the emitted parton. A very striking difference between  \pythia -\pythiaPB\  and \pythia\  is observed:  \pythia -\pythiaPB\ applies angular ordering, resulting in the relation $\qt = (1-z) q' $ while \pythia\ uses \pt -ordering with the relation $\qt \simeq q'$ (including corrections from kinematic constraints). The relation $\qt \simeq q'$ is clearly seen in Fig.~\ref{fig:ISR_zdistribution}: at larger $q'$ emissions with small \qt\ are absent.
Thus the absence of soft gluon emissions in the standard \pythia\ initial state parton shower comes from the \pt -ordering condition, which also implies significant constraints on \zM . As explicitly shown in Ref.~\cite{Mendizabal:2023mel}, soft gluons are important at parton level, but play a smaller role on hadron level, as they act only as \pt -kinks in the string in the Lund hadronization model.

\clearpage

\bibliographystyle{mybibstyle-new.bst}
\raggedright  
\providecommand{\href}[2]{#2}\begingroup\raggedright\endgroup

\end{document}